\documentclass{IEEEtran}
\usepackage{amsmath,amssymb,amsfonts}
\usepackage{graphicx}
\usepackage[bookmarks=false]{hyperref}
\usepackage{siunitx}
\usepackage{cite}
\usepackage{textcomp,nicefrac}
\usepackage{color}
\usepackage{url}
\usepackage{gensymb}
\usepackage{color,soul}
\usepackage{cleveref}
	\crefname{figure}{Fig.}{Figs.}
	\Crefname{figure}{Fig.}{Figs.}
	\crefname{table}{Table}{Tables}
	\Crefname{table}{Table}{Tables}
\usepackage{multirow}
\usepackage{makecell}
\definecolor{purple}{rgb}{0.5,0.1,1}

\def\BibTeX{{\rm B\kern-.05em{\sc i\kern-.025em b}\kern-.08em
T\kern-.1667em\lower.7ex\hbox{E}\kern-.125emX}}
\begin{document}

\title{Protocols for Healing Radiation-Damaged Single-Photon Detectors Suitable for Space Environment}
\author{Joanna Krynski, Nigar Sultana, Youn Seok Lee, Vadim Makarov, and Thomas Jennewein
\thanks{Nigar Sultana and Thomas Jennewein are currently with the Department of Physics and Astronomy and Institute for Quantum Computing, University of Waterloo, Waterloo, Ontario, N2L~3G1, Canada.

Joanna Krynski, Youn Seok Lee and Vadim Makarov were with the Department of Physics and Astronomy and Institute for Quantum Computing, University of Waterloo, Waterloo, Ontario, N2L~3G1, Canada (email:jkrynski@uwaterloo.ca).}}

\maketitle

\begin{abstract}
Single-photon avalanche detectors (SPADs) are well-suited for satellite-based quantum communication because of their advantageous operating characteristics as well as their relatively straightforward and robust integration into satellite payloads. However, space-borne SPADs will encounter damage from space radiation, which usually manifests itself in the form of elevated dark counts. Methods for mitigating this radiation damage have been previously explored, such as thermal and optical (laser) annealing. Here we investigate in a lab, using a CubeSat payload, laser annealing protocols in terms of annealing laser power and annealing duration, for their possible later use in orbit. Four Si SPADs (Excelitas SLiK) irradiated to an equivalent of 10 years in low Earth orbit exhibit very high dark count rates ($>$300~kcps at $-$22~$\celsius$ operating temperature) and significant saturation effects. We show that annealing them with optical power between 1 and 2~W yields reduction in dark count rate by a factor of up to 48, as well as regaining SPAD sensitivity to a very faint optical signal (on the order of single photon) and alleviation of saturation effects. Our results suggest that an annealing duration as short as 10~s can reduce dark counts, which can be beneficial for power-limited small-satellite quantum communication missions. Overall, annealing power appears to be more critical than annealing duration and number of annealing exposures.
\end{abstract}


\section{Introduction}
\label{sec:introduction}

\IEEEPARstart{S}{ingle-photon} avalanche photodiodes (SPADs) are widely used in various fields, including applications in remote sensing and LIDAR \cite{krainak2010}, medical imaging \cite{Palubiak2011,Mu2015}, and classical communications \cite{steindl2017}. SPADs have also been extensively used in space, where their ease of integration and operation, as well as wide spectral sensitivity make them preferable over other single-photon detectors such as photomultiplier tubes and superconducting nanowires \cite{hadfield2009}. SPADs will be particularly integral in the facilitation of a quantum communication systems as their high detection efficiency, low timing jitter, low dark count rate and low afterpulsing probability make them prime candidates for satellite-based quantum receivers, as demonstrated by the Micius satellite and missions to be launched in the near future \cite{kim2011,bourgoin2012,yang2019,jennewein2014,tan2013}.

One dominant issue is the impact of space radiation, which degrades detector performance, particularly in increasing the noise associated with thermal generation of carriers, a phenomenon known as dark count rate (DCR) \cite{sun1997,tan2013}. Elevated DCR originates from the introduction of defects in the semiconductor lattice produced by interactions with particulate radiation, particularly low-energy protons \cite{srour2003}. Accumulation of these defects and, consequently, steadily increasing DCR pose major constraints on the lifespan of SPADs within quantum satellite payloads. One study found that a quantum communication satellite in low-Earth orbit (LEO) could become unusable within several weeks, and even this estimate does not account for unpredictable solar proton events that can deposit large amounts of defects at once \cite{anisimova2017}. This issue is critical for quantum communication applications where a maintained dark count rate as low as $200$~counts-per-second (cps) is required to facilitate a high-fidelity transmission \cite{bourgoin2012}. Therefore, in order to prolong the usable lifetime of quantum missions, methods of counteracting this radiation damage will be essential in satellites hosting SPADs. 

At a fundamental level, reduction of DCR amounts to removing or re-arranging the radiation-induced defects. This can be achieved via annealing, which has been used for decades in the semiconductor industry in the context of improving material characteristics, such as electrical conductivity \cite{srour2003,boyd1980}. Thermal annealing via a hot-flow oven or thermoelectric (TEC) module as a method of controlling the radiation-induced DCR in SPADs has been investigated and successfully demonstrated \cite{anisimova2017,dsouza2021,yang2019}. The method of using TECs is feasible for satellite missions and even then requires heating to near $100~\celsius$, which risks damaging the TEC, and necessitates a high payload power allowance. An alternative approach to thermal annealing is laser (optical) annealing. Both pulsed and continuous-wave lasers can be used to facilitate localized melting and re-crystallization of the lattice, with the energy and pulse length dictating the magnitude and depth of energy deposition in the material \cite{sorensen1981}. Laser annealing is advantageous because its highly localized energy deposition does not risk degradation of the material or redistribution of imperfections, and the recovery period is much shorter as compared with traditional thermal annealing \cite{khaibullin1978}. With respect to laser annealing SPADs, Bugge et al.\ \cite{bugge2014} observed a decrease in dark counts after exposing SPADs to bright laser beam, and Lim \cite{lim2017} showed a similar behaviour when deliberately annealing irradiated devices with a free-space coupled laser of optical power on the order of a few watt. Moreover, Lim et al.\ \cite{lim2017} showed that laser annealing was quicker in achieving greater reduction in DCR compared to thermal annealing since energy is not wasted in heating the whole SPAD enclosure but rather delivers highly localized uniform heating straight to the active area. Real-time in-orbit laser annealing of SPADs impinged by space radiation has yet to be demonstrated. The Cool Annealing Payload Satellite (CAPSat) which was deployed from the ISS on October~12, 2021 was intended to study this technique in orbit \cite{CAPSatPressRelease}. Another mission set to launch soon, the Satellite Entanglement Annealing QUantum Experiment (SEAQUE), will employ laser annealing throughout its lifetime \cite{ortiz2022}. To this date, no periodic laser annealing in the space environment, especially on a small satellite bus with limited thermal dissipation and restricted power budget, has been shown.

In order to study the viability of this technique in the context of a small satellite payload, we present a ground study of three protocols for laser annealing in a vacuum environment using a detector module designed for in-orbit annealing \cite{sultana2022}. The goal of this study is to determine the operational range of this technique, with particular focus on choice of laser power and annealing duration, as well as heat generation during the annealing process. The four SPADs to be annealed were previously irradiated, and are integrated into an electrical module designed under size, mass and power constraints for the CAPSat mission. In contrast to previous tests \cite{lim2017}, the SPADs are directly fiber-coupled to the annealing laser, in line with the detector module configuration in CAPSat and SEAQUE. Fiber-coupling in space-based modules is advantageous compared with free-space configurations because it eliminates the risk of the annealing beam misaligning from the active area during launch, as well as provides a more concentrated exposure onto the active area. Because of prior exposure to radiation \cite{anisimova2017}, the SPADs possess very high initial dark count rates ($>300$~kcps) and significant saturation effects. We anneal them with powers ranging from $1$~W to over $2$~W and for annealing periods of up to $16$~min. We report up to a $48$-fold reduction of DCR as well as evidence of improvement in signal-to-noise ratio and reduction in saturation.

\section{Experimental setup}

The four SPAD devices to be laser annealed are silicon single-photon APDs (Excelitas SLiK, $180~\micro$m diameter active area), packaged with a two-stage TEC and fiber connectors intended to focus the beam onto the active area through a small lens. The SPADs were irradiated with $105$~MeV protons in 2017 at the TRIUMF Proton Irradiation facility \cite{dsouza2021}. The maximal cumulative fluence achieved ($2 \times 10^{10}$ protons/cm$^2$) was equivalent to $10.5$~years in a $600$-km polar orbit behind $10$~mm of aluminum shielding, as determined by calculations using the SPENVIS radiation modelling package \cite{spenvis}. Additionally, these SPADs were repeatedly thermally annealed for one hour from their nominal temperature of $-78~\celsius$ to either $-40~\celsius$ or $+80~\celsius$ by running the in-built TEC in reverse \cite{dsouza2021}. Annealing occurred in one set after each round of irradiation, while in the other set, it occurred after a dark count rate threshold of $2$~kcps was exceeded. DSouza et al.\ \cite{dsouza2021} calculated that the effective energy of the protons after shielding is $92.8$~MeV \cite{dsouza2021}. Using non-ionizing energy loss (NIEL) damage curve generated in SPENVIS for damage equivalent proton energy of $105$~MeV with a relative degradation per unit NIEL of $1 \times 10^{-11}$~g(Si)$\,$MeV$^{-1}$ \cite{jun2003}, we calculate that the NIEL after the shielding is $2.7 \times 10^{-3}$~MeV$\,$cm$^2$g$^{-1}$. From this one can estimate the displacement damage dose (DDD) incurred over the course of the irradiation \cite{summers1993}
\begin{equation}
\label{eq:DDD}
\text{DDD} = \int_{E_0}^{E} \Phi(E) \text{NIEL}(E) dE .
\end{equation}
Because the beam is monoenergetic, this integral reduces to the product of the two quantities, yielding a displacement damage dose (DDD) of $5.95 \times 10^{7}$~MeV$\,$g$^{-1}$ in our case. This is useful in comparing the damage incurred between energies of one particle type and between particle types (protons, neutrons, electrons).

\begin{figure}
	\centering
	\includegraphics[width=\columnwidth]{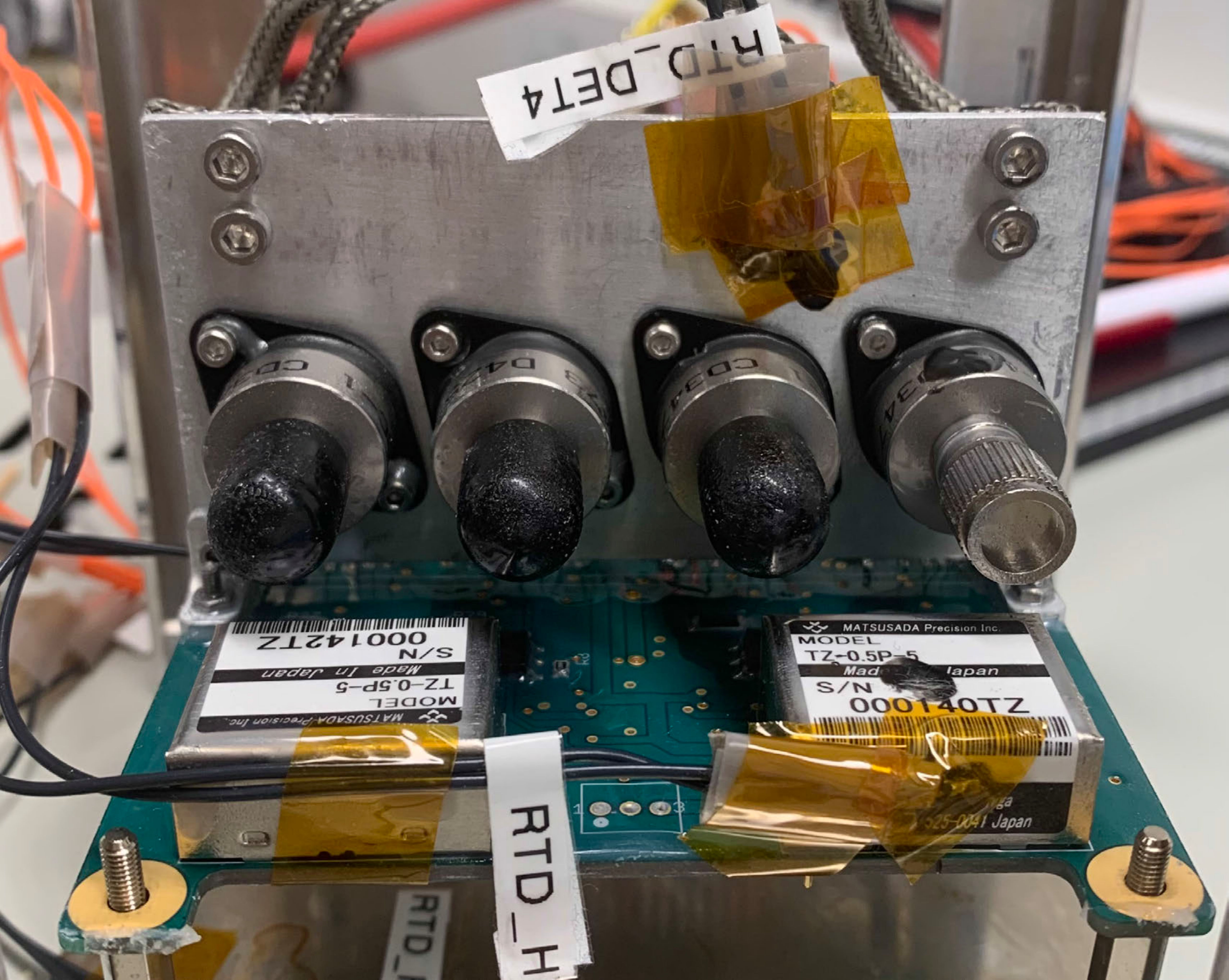}
	\caption{Detector module (DM) with four capped Si single-photon avalanche diodes (SPADs) atop a printed-circuit board providing temperature control and bias voltage.}
	\label{fig:CAPSAT}
\end{figure}

The devices were integrated into an engineering model of the detector module (\cref{fig:CAPSAT}). The four devices are mounted to an aluminum bracket side-by-side, and attached to a printed circuit board (PCB) that contains circuitry for detector bias, passive quenching and readout, as well as temperature control. The DM is controlled via an externally connected embedded programmable system-on-chip Cypress PSOC3 development kit, which also runs a thermal control loop. Detailed description of the CubeSat module can be found in \cite{sultana2020,sultana2022}.

\begin{figure}
	\centering
	\includegraphics[width=0.96\columnwidth]{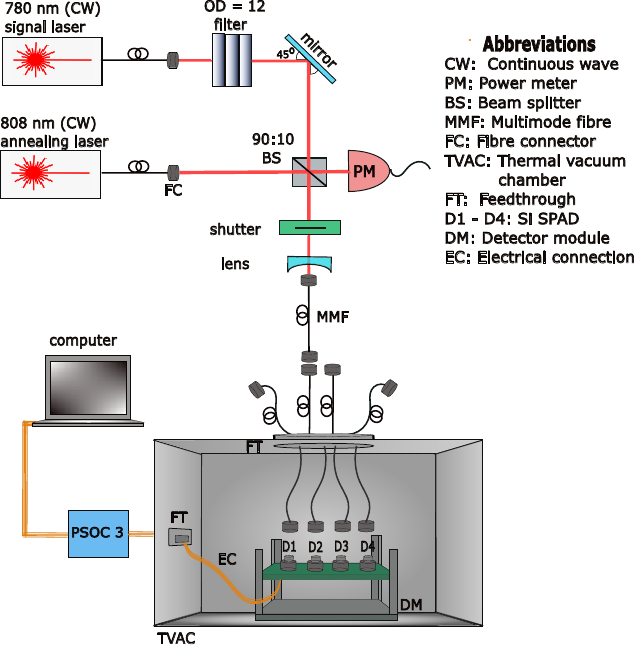}
	\caption{Schematic representation of test set-up. Details in text.}
	\label{fig:optical set-up}
\end{figure}

\begin{figure}
	\centering
	\includegraphics[width=0.65\columnwidth]{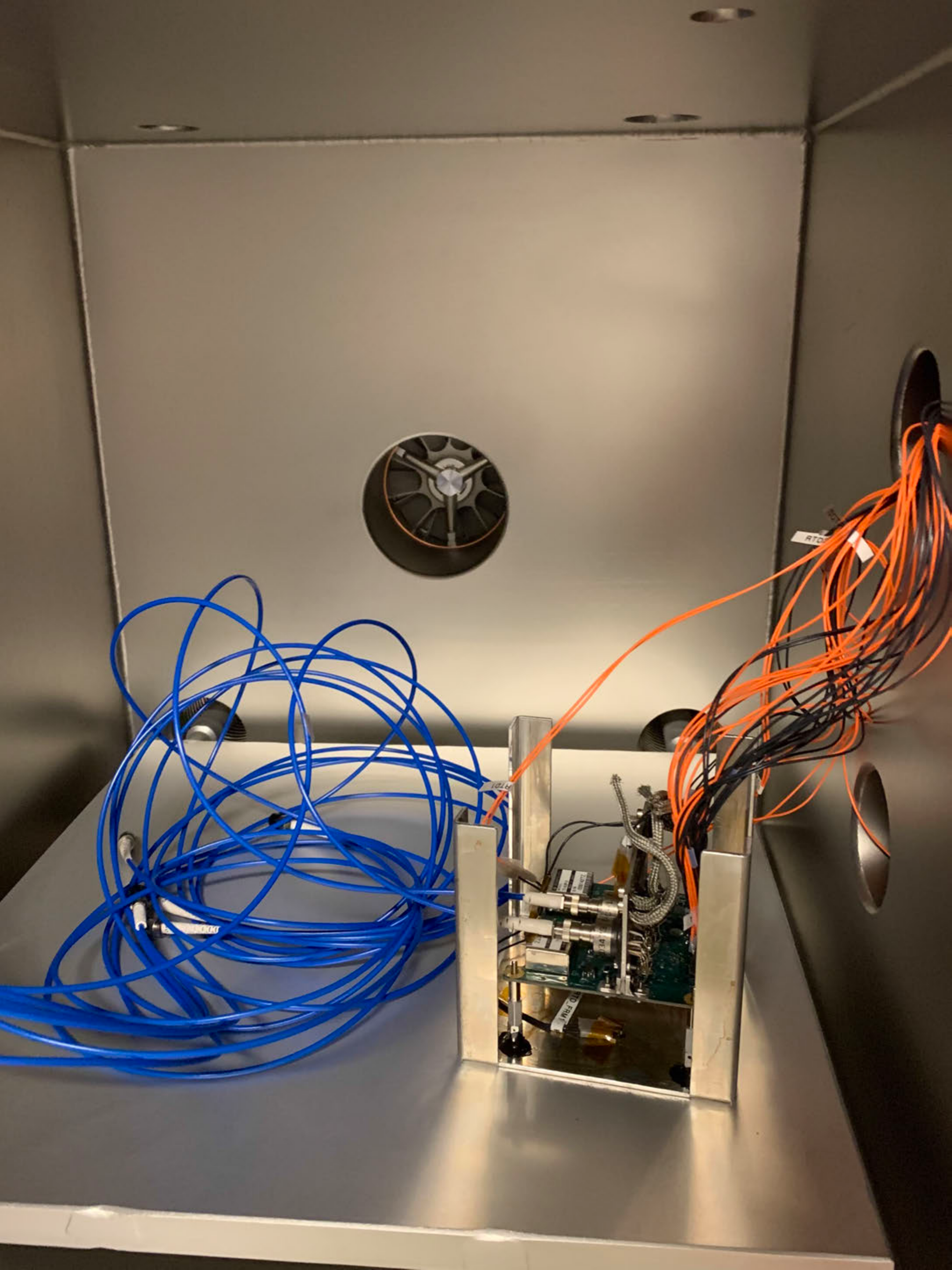}
	\caption{Detector module inside a thermal vacuum chamber (TVAC). Blue cables are vacuum-suitable optical fibers, while the orange and black wires are the electrical cables connected to the microcontroller situated outside the vacuum chamber.}
	\label{fig:TVAC}
\end{figure}

The test set-up, shown in \cref{fig:optical set-up}, is comprised of bench-top optics and the PSOC3 kit at ambient pressure, and the DM inside a thermal vacuum chamber (TVAC) at an average pressure of $10^{-6}$~torr ($\sim 0.001$~atm; see \cref{fig:TVAC}). The optical input is comprised of a tunable high power multi-mode (MM) $808$~nm laser diode (Jenoptik JOLD-30-DC-12) used for annealing and a $780$~nm laser diode (Toptica DL PRO 780 FD2) used to test SPAD sensitivity. Henceforth these will be called the annealing and the signal laser, respectively. The annealing laser is chosen because of its powerful output (up to $30$~W) and central wavelength well-within the absorption range of the SPAD, while the signal laser wavelength is chosen as it is close to the peak sensitivity of the SPAD. The fiber-coupled annealing laser is collimated by a lens, after which it enters a $90\!:\!10$ beamsplitter (BS); $10\%$ of the beam transmits through to a power meter (Thorlabs S121C with PM120) and $90\%$ of the beam reflects towards a fast electric-shutter (Thorlabs SH05 with Thorlabs SC10 controller) where it is focused by a lens into a multi-mode fiber (MMF). The signal laser is initially fiber coupled to free-space and passes through neutral density filters (maximum optical density, $\text{OD} = 12$) to reach single-photon level. This signal beam is injected into the other port of the BS, and couples with the annealing laser beam in the MMF. Because the fiber termination inside the TVAC would be inaccessible during operation, the power of the signal laser at the SPAD active area is estimated using the power measured at the $90\%$ output port of the BS. The multi-mode fiber enters the TVAC via a fiber feed-through and connects to one of the SPADs. The detector module electrical connections are relayed with vacuum-suitable cables and are connected to the PSOC via another vacuum electrical feed-through. Three resistance temperature detectors (RTD; PT-1000 type) are placed on the aluminum mounting plate, on the SPAD high-voltage supply, and in the center of the PCB, to monitor the local heating of the respective areas. The RTDs' wires also exit the TVAC through the electrical feed-through and their resistance values are measured using multimeters (Fluke 289). DM operation is controlled using a PSOC3 microcontroller and collected data is sent to a computer via serial communication. The data collected includes those related to DM operational status (such as TEC current) as well as those related to SPAD readout (such as dark count rate). 
 
\section{Methods}

\subsection{Laser annealing protocols}

Three laser annealing protocols are tested: \emph{(a)} increasing annealing power with a fixed exposure of three minutes; \emph{(b)} fixed annealing power with a single exposure of increasing duration; and \emph{(c)} fixed annealing power with three exposures of the same duration. The laser powers for the latter two protocols were chosen based on the results from the first protocol. The number of exposures is increased to three for the third protocol to investigate if repeated exposures of the same energy (i.e.,\ same power and duration) provide additional insight into the defect removal mechanism. SPAD characterization occurs after each annealing exposure. \Cref{Table:annealingparam} summarizes the selected annealing powers and duration of exposures for each of device.

The optical annealing power listed in \cref{Table:annealingparam} is estimated at the end of the MMF that connects directly to the SPAD package. It is calculated via a pre-calibrated ratio between the MMF and PM.

For each round, the pump current of the annealing laser is monitored to achieve the correct annealing power, after which the fast shutter is opened for a pre-determined exposure duration, and the SPAD is annealed. The detector module is not powered during annealing and at least for $30$~s afterwards, to allow for the DM to cool. The resistance of the RTDs is measured during the exposure, giving an insight into the heating process of the area surrounding the annealed detector.

\subsection{Characterization}

Detector characterization is conducted in complete darkness, with a blackout curtain covering the viewport of the TVAC to block out background light. The PSOC program controls which single SPAD will be biased and its temperature. Measurements are taken at three SPAD temperatures: $-22$, $-10$, and $0~\celsius$. The temperature settling time is approximately $60$~s, after which the SPAD is biased, and the breakdown voltage is determined by observing when the readout program begins to register counts, which will be lower than the true value due to the discriminator threshold. For each temperature, dark counts from the selected SPAD are logged once per second for one minute while the laser annealing beam is off and while the shutter is closed. In a similar fashion, for each temperature, detection counts from the selected SPAD are logged while the shutter is open and the SPAD active area is exposed to the signal laser. Additional characterization of the SPADs, such as measurements timing jitter or recharge time, are not possible in our set-up.

\begin{table}
\centering
\caption{Annealing exposure parameters}
\resizebox{\columnwidth}{!}{%
\begin{tabular}{@{\extracolsep{1.8ex}}c@{}c@{}c@{}c@{}c@{}} 
\hline\hline
SPAD & \makecell{Annealing\\ power (W)} & \makecell{Number of\\ exposures} & \makecell{Maximum\\ duration (min)} & \makecell{Initial DCR (kcps)\\ at $-22~\celsius$} \\ 
\hline
$1$	& $0$--$2.3$	& $1$	& $3$		& $325$	\\
$2$	& $1.5$				& $1$	& $14$	& $404$	\\
$3$	& $1.2$				& $3$	& $16$	& $414$	\\
$4$	& $1.8$				& $3$	& $16$	& $392$	\\
\hline\hline
\end{tabular}
}
\label{Table:annealingparam}
\end{table}

\begin{table}
\centering
\caption{Thermal data}
\label{tab:temperature-increase}
\resizebox{\columnwidth}{!}{%
\begin{tabular}{@{\extracolsep{1.8ex}}c@{}c@{}c@{}c@{}c@{}}
\hline\hline
\multirow{2}{*}{\makecell{Peak\\ power\\ (W)}} & \multirow{2}{*}{\makecell{Estimated peak\\ temperature ($\celsius$)\\ (extracted from \cite{lim2017})}} & \multicolumn{3}{@{}c@{}}{Maximum $\Delta \text{T}$ ($\celsius$) during exposure} \\ \cline{3-5} 
& & \multicolumn{1}{c}{PCB} & \multicolumn{1}{c}{\makecell{Aluminum\\ mount}} & HV supply \\
\hline
$2.3$	& $>160$	& $4.69$	& $2.65$	& $3.89$ \\
$1.5$	& $120$		& $2.26$	& $3.68$	& $3.04$ \\
$1.2$	& $110$		& $2.53$	& $3.56$	& $3.14$ \\
$1.8$	& $135$		& $3.27$	& $4.56$	& $4.12$ \\
\hline\hline
\end{tabular}%
}
\end{table}

\subsection{Operational voltage and saturation}

Our detector employs a simple passive-quenching circuit, which has a relatively low saturation count rate but is entirely adequate for the high-loss ground-to-satellite channel \cite{kim2011,bourgoin2012,jennewein2014}. In the passive-quenching circuit, the SPAD's deadtime is not well-defined and tends to increase with the bias voltage applied, owing to avalanches failing to quench quickly at higher bias voltages \cite{kim2011}. The saturation count rate is thus lower at a higher bias voltage, while both the probability of dark count occurring (when the SPAD is not in the deadtime) and photon detection efficiency increase with applied bias \cite{migdall2013}. Owing to the random nature of avalanche quenching and bias recovery, this saturation effect cannot be well corrected mathematically.

\begin{figure}
	\centering
	\includegraphics[width=0.9\columnwidth]{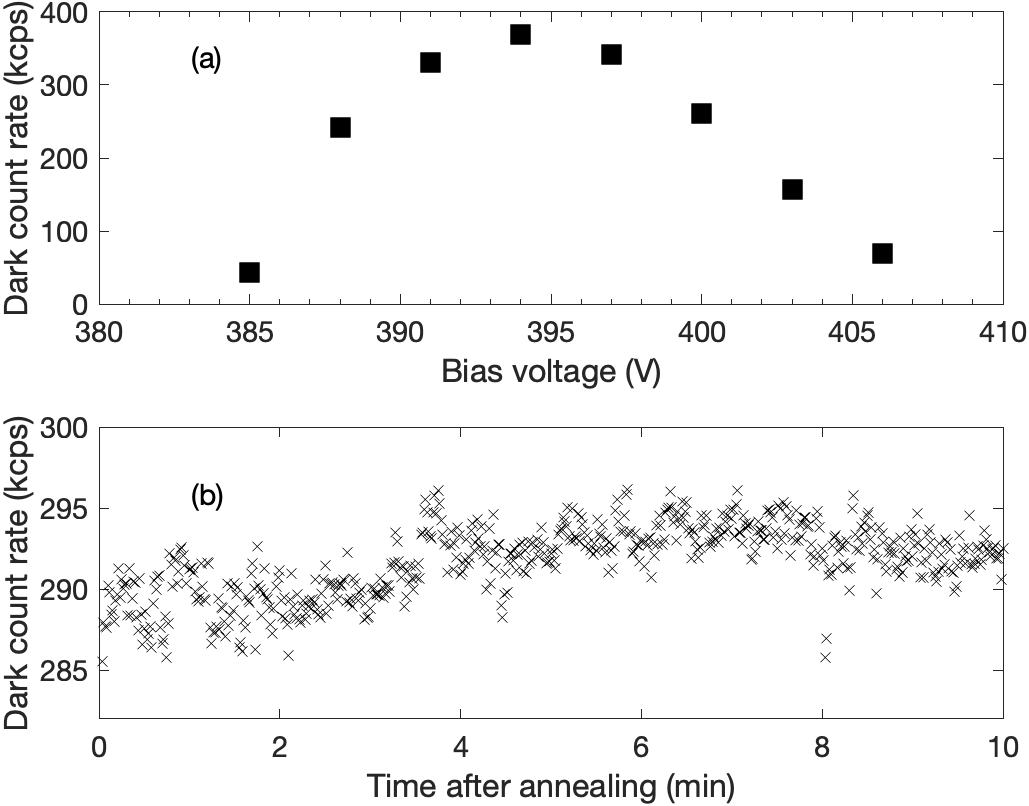}
	\caption{(a)~Observed dark count rate (DCR) as function of SPAD~1 bias voltage. Saturation point occurs at about $9$~V above breakdown, therefore, the SPAD should be operated at a bias lower than this to ensure reliable recording of effective DCR. (b)~Dark count rate of SPAD~1 measured for $10$~min after the conclusion of a round of annealing at $1.5$~W for three minutes. No notable spikes in DCR appear during this period.}
	\label{fig:saturated-voltage-scan}
\end{figure}

Unfortunately, due to the very high initial DCR of our irradiated SPADs ($\sim 300$~kcps), saturation already occurs between $9$--$12$~V above breakdown (\cref{fig:saturated-voltage-scan}a). Below that voltage, DCR rises with bias and the detector is not saturated. We thus perform our measurements at $6$~V above breakdown, unless otherwise noted, to be sufficiently below saturation but still with a reasonably good detection efficiency. In an actual satellite operation, DCR will stay lower than in our tests, and the SPADs might be biased about $20$~V above breakdown, to increase their detection efficiency slightly \cite{anisimova2017,dsouza2021,lim2017}.

\begin{figure}
	\includegraphics[width=0.95\columnwidth]{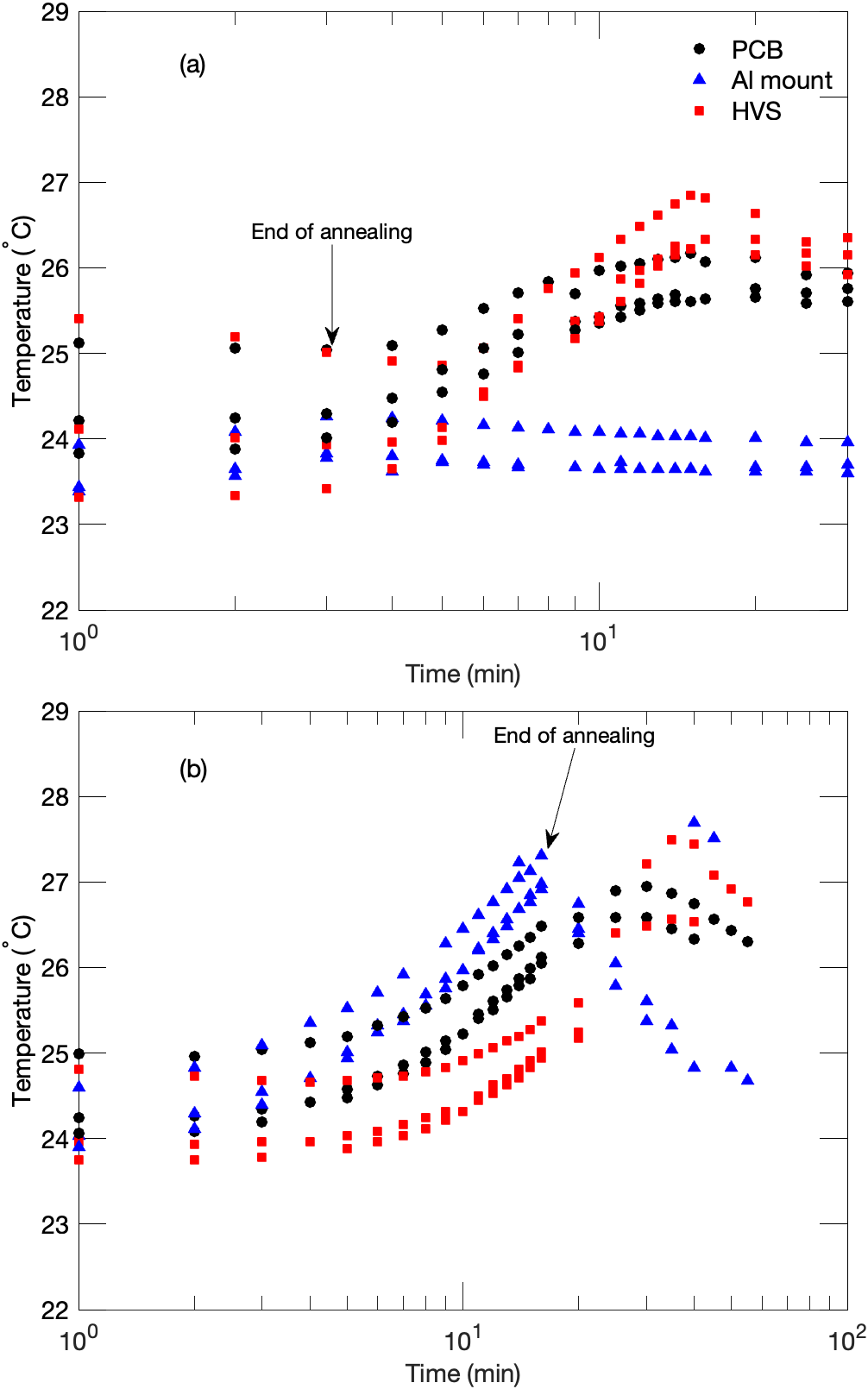}
	\caption{Temperature data recorded for (a)~$3$~min at $0.95$--$1.1$~W and (b)~$14$~min annealing at $1.2$~W, as well as for the subsequent module operation. RTDs are placed in three locations: in the centre of the module’s electronics (PCB); as close as possible to the detectors on the aluminum mounting bracket (Al mount) and on the module's high-voltage supplies (HVS).}
	\label{fig:temperature-data}
\end{figure}

\section{Results}

\subsection{Thermal vacuum operation}

While we are unable to measure the temperature directly at the SPAD active area in our set-up, Lim et al.\ \cite{lim2017} reported in their similar set-up that the peak temperature measured by the active area thermistor is reached within $60$~s of the exposure, with temperature reaching above $90~\celsius$ for optical powers greater than $1$~W. \Cref{tab:temperature-increase} shows the estimated peak temperature for the peak optical powers in the current study, using the data from \cite{lim2017}. It is expected that even greater temperatures are reached in our set-up, as the SPADs are directly fiber-coupled to the annealing laser. 

The vacuum pressure remained fairly constant, with no significant outgassing events observed. There were no issues reported in operating in vacuum with respect to thermal dissipation. The CubeSat module chassis acted as an excellent thermal conductor between the electronics and the chamber plate, to which the chassis was fixed. During annealing and subsequent module operation, the maximum temperature difference recorded by the RTDs was approximately $5~\celsius$ above ambient (\cref{tab:temperature-increase}). \Cref{fig:temperature-data}a shows that, for short annealing exposures, the greatest contributor of temperature rise was the regular operation of electronics, because the largest increase in temperature is seen by the detector after annealing is complete and the module is turned on. However, for long annealing exposures, shown in \cref{fig:temperature-data}b, the temperature increases during annealing, especially in the aluminum mount holding the detectors. After annealing is complete, the aluminum mount temperature drops rapidly.

\subsection{SPAD~1: Post-annealing dark count rate}

The DCR of SPAD~1 is observed for $10$~min immediately after each of the $20$ annealing rounds to investigate if there is any residual change (e.g.,\ temperature variation) in the DM that could contribute to additional thermal counts. \Cref{fig:saturated-voltage-scan}b shows typical data with no significant variations or jumps in DCR after annealing, confirming that the heating of the active area is present only when the beam is incident and cooling occurs rapidly after the beam is shut off.

\begin{figure}
	\hspace{0.4mm}\includegraphics[width=0.922\columnwidth]{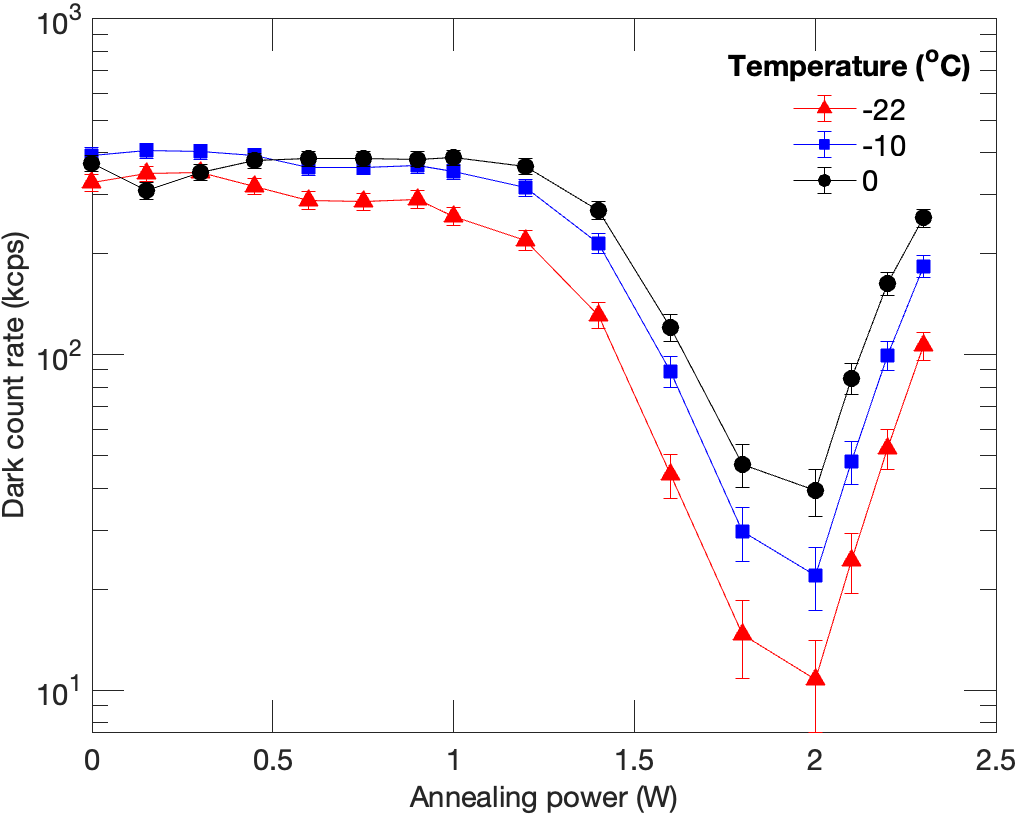}
	\caption{Observed DCR of SPAD~1 measured at three operating temperatures after annealing at various powers for three minutes.}
	\label{fig:variable-annealing-power-DCR-temp}
\end{figure} 

\begin{figure}
	\includegraphics[width=\columnwidth]{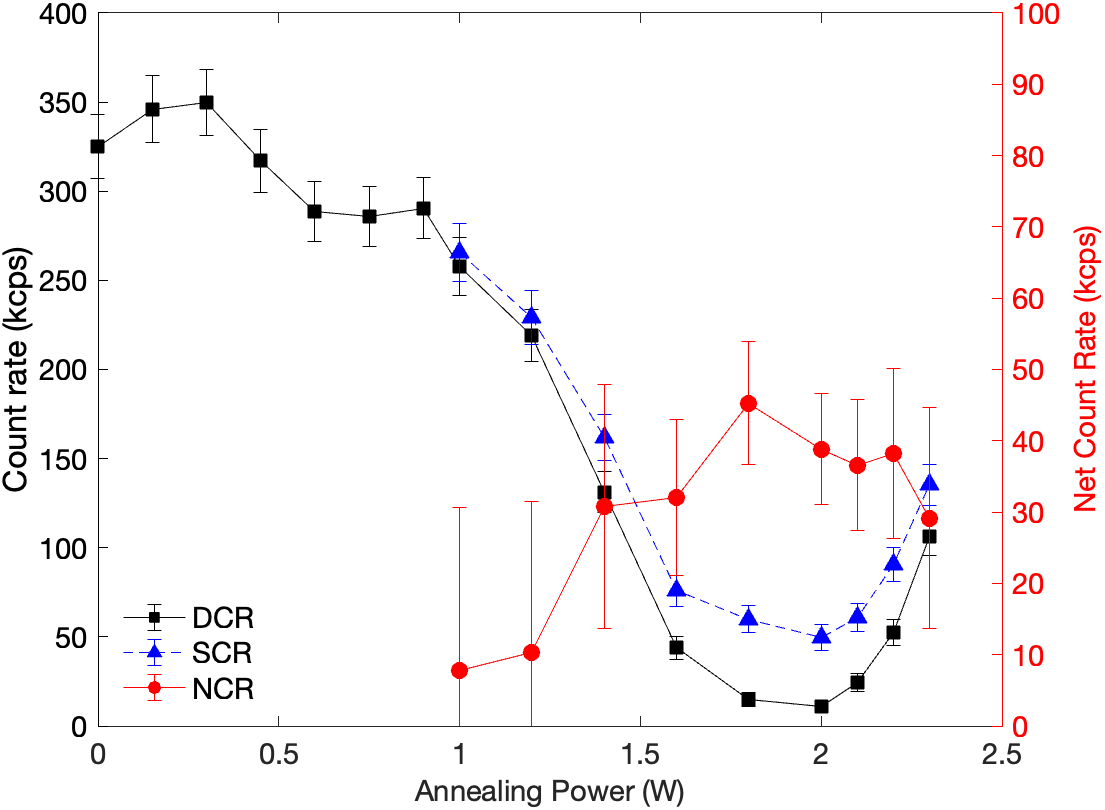}
	\caption{Count rates measured when signal laser is off (DCR), on (signal count rate; SCR), and the difference between the two count rates (net count rate; NCR) for SPAD~1 at an operating temperature of $-22~\celsius$.}
	\label{fig:variable-power-DCR-signal-NCR}
\end{figure}

\subsection{SPAD~1: Fixed duration, variable annealing power}

Annealing SPAD~1 with laser powers below $1$~W for a fixed three-minute exposure does not yield significant decrease in DCR (\cref{fig:variable-annealing-power-DCR-temp}). The dark count rate decreases rapidly for powers between $1$~W and $2$~W, but increases again with all subsequent higher-power exposures. This increase likely indicates that permanent damage is being done to the SPAD active area. With a pre-annealing DCR of $325$~kcps (at $-22~\celsius$) and lowest post-annealing DCR of $10.8$~kcps, a dark count reduction factor (DCRF) of about $30 \pm 9.3$ is observed.

Before annealing, SPAD~1 count rates while exposed to the signal laser (signal count rate, or SCR) lead to detector saturation, that is, the observed count rate drops below the DCR; after annealing with $1$~W, however, we find that the SPAD no longer saturates when illuminated with the signal laser. As the annealing power increases, the signal counts become more distinguishable from noise and the net count rate (difference between SCR and DCR; NCR) increases (\cref{fig:variable-power-DCR-signal-NCR}). Above $1.8$~W of annealing power, the NCR begins to drop, as DCR increases again.

There are no noticeable improvements to the DCR for exposures of less than $1$~W in accordance with the results of Lim et al.\ \cite{lim2017}, despite the use of fiber-coupling, which concentrates the beam on the active area. This result could suggest that there is a threshold beam energy (combination of laser power and exposure duration) required to begin the annealing process. That said, it is possible that there is indeed some annealing occurring, but saturation effects prevent an accurate measurement of count rates with the signal laser. This would imply that the reduction in true DCR is even greater from the observed DCR, and our results thus provide an lower bound to the achievable improvement.

\subsection{SPADs~2, 3, 4: Fixed power, variable annealing duration}

\begin{figure}
	\includegraphics[width=0.95\columnwidth]{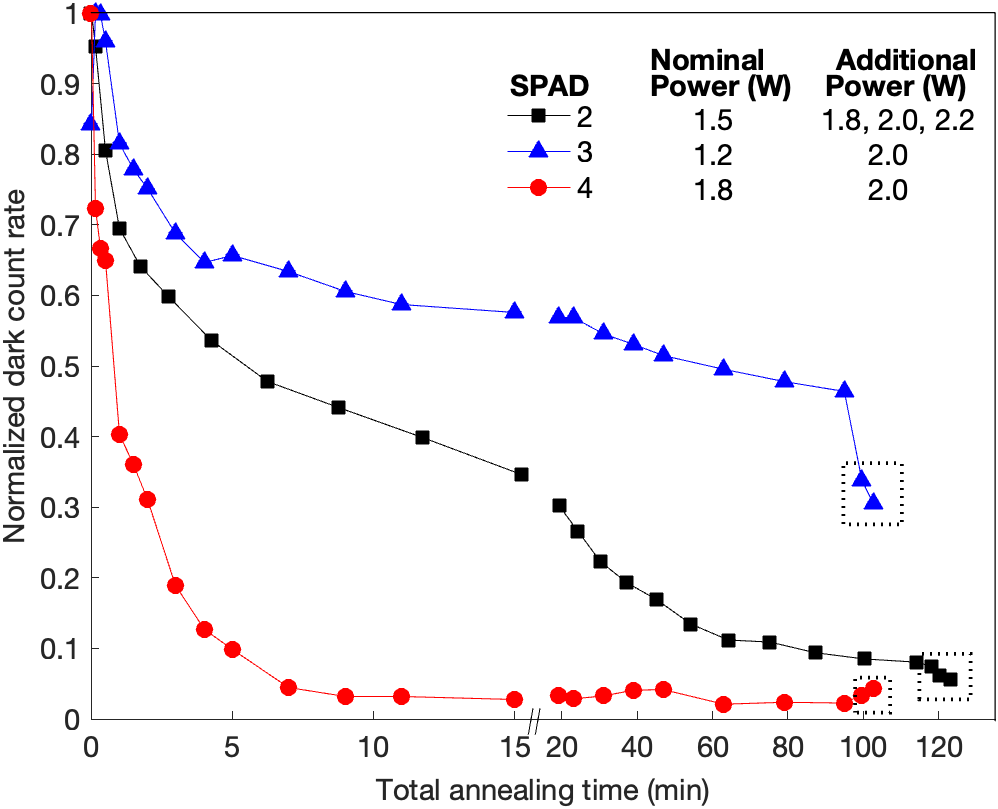}
	\caption{Normalized DCR (relative to the pre-annealing rate) as a function of total annealing time, which is the total time the annealing laser is incident on the active area. Additional points of shorter rounds of annealing at higher powers are indicated by dotted boxes. Error bars are too small to be seen at this scale.}
	\label{fig:variable-time-DCR-normalized}
\end{figure}

Because noticeable increases in NCR were seen in SPAD~1 for powers above $1$~W, each remaining SPAD is annealed at a fixed higher power (SPAD~2 at $1.5$~W, SPAD~3 at $1.2$~W, and SPAD~4 at $1.8$~W), but with variable annealing duration. Normalizing the results with respect to each SPAD's maximum recorded DCR shows that higher annealing power yields a steeper drop in DCR with the most drastic decrease after the first and shortest exposures of $10$~s (\cref{fig:variable-time-DCR-normalized}). It does not appear that annealing for longer periods guarantees greater reduction in DCR. A plateau in the DCR is observed in all three SPADs for long exposure duration. The DCRF for $1.2$, $1.5$, and $1.8$~W annealing tests is $1.8 \pm 0.15$, $13 \pm 2.2$, and $48 \pm 17$, respectively. To confirm the plateaus are indeed the lowest achievable DCR for each annealing power, the SPADs are annealed with higher laser powers for another three minutes. Additional annealing up to $2$~W of SPADs~2 and 3 further reduced their DCR, with the DCRF increasing to $2.8 \pm 0.008$ and $18 \pm 0.12$, respectively. For SPAD~4, which was prior annealed at $1.8$~W, the DCR actually rose after additional $2$~W exposure. Accordingly, the DCRF for this SPAD reduced to $24 \pm 0.19$. Although on average the DCR decreased in this SPAD, the DCR did not monotonically decrease as in the other two SPADs. This likely means that annealing at this power level may actually introduce additional defects in the crystal lattice and damage the detector's sensitive area prior to the final exposure at $2$~W. 

\begin{figure}
	\includegraphics[width=0.95\columnwidth]{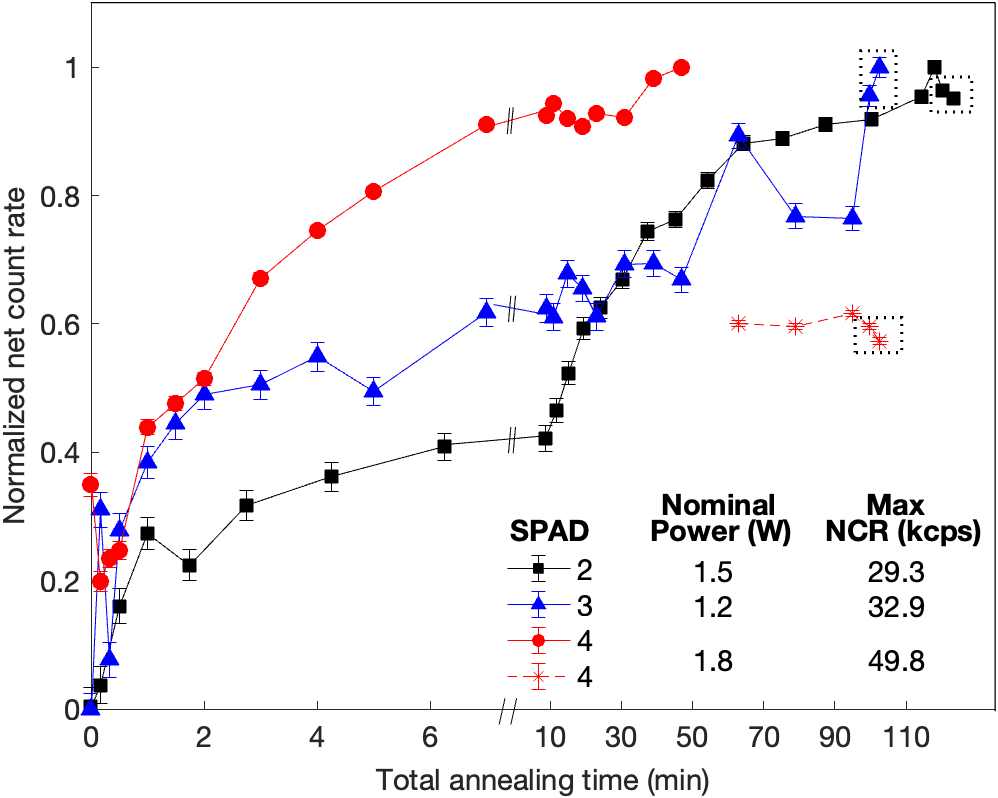}
	\caption{Observed net count rate after normalization with respect to pre-annealing count rate, for the SPADs~2, 3, and 4 annealed at various powers and for various duration. Additional points of shorter rounds of annealing at higher powers are indicated by dotted boxes. Dotted star points for SPAD~4 are taken at lower bias voltage.}
	\label{fig:variable-time-NCR-normalized}
\end{figure}

\Cref{fig:variable-time-NCR-normalized} shows the NCR normalized with respect to the pre-annealing NCR of the SPADs annealed with single annealing power. Net count rate for SPADs~2, 3, and 4 increased to a maximal rate of $29.3$, $32.9$, and $49.8$~kcps, respectively, showing a sizeable gap between measured signal and dark counts. However, we see that NCR drops for SPAD~3 and SPAD~4 in the last two fixed power exposures even before $2$~W is applied in addition to the previous $1.2$ and $1.8$~W. It was not possible to determine the reason for the drop in NCR for SPAD~3; however, for SPAD~4, it was determined afterwards that the excess voltage was reduced below the voltage used prior (from $6$~V excess to $3$~V excess). Despite being at this lower excess bias, the additional burst of higher power annealing reduced the NCR for SPAD~4 (red stars in dotted box). SPAD~2 similarly suffered a reduction in NCR after a monotonically increasing behavior prior to the additional exposures (black squares in dotted box). It is conceivable that the extra $2$-W exposures introduced additional defects, rather than remove them.

\subsection{Repeated versus single annealing exposure}

Exposing SPADs~3 and 4 repeatedly to the same energy (i.e.,\ the same laser power for the same duration) did not result in faster or higher reduction in DCR compared with SPAD~2, which was exposed only once. The rate at which DCR diminished appears to be solely dependent on the laser power, as the DCRF for SPADs~2, 3, and 4 was higher for a higher annealing power applied. This is analogous to higher temperature annealing with convective heating (oven, TEC, etc.). It is well known that annealing at higher temperatures leads to more rapid recovery towards nominal dark current levels (for example, \cite{eremin1997,becker2003}). According to our thermal data, presented in \cref{tab:temperature-increase}, the peak temperatures recorded by the temperature sensors closest to the detector active area occurred in descending order with respect to annealing power, with $1.8$-W annealing leading to the highest peak temperature followed by the $1.5$-W and finally $1.2$-W annealing. Therefore, it is expected that a higher annealing power is analogous to higher annealing temperature, which would increase the rate of defect removal. 

\begin{figure}
	\centering
	\includegraphics[width = \columnwidth]{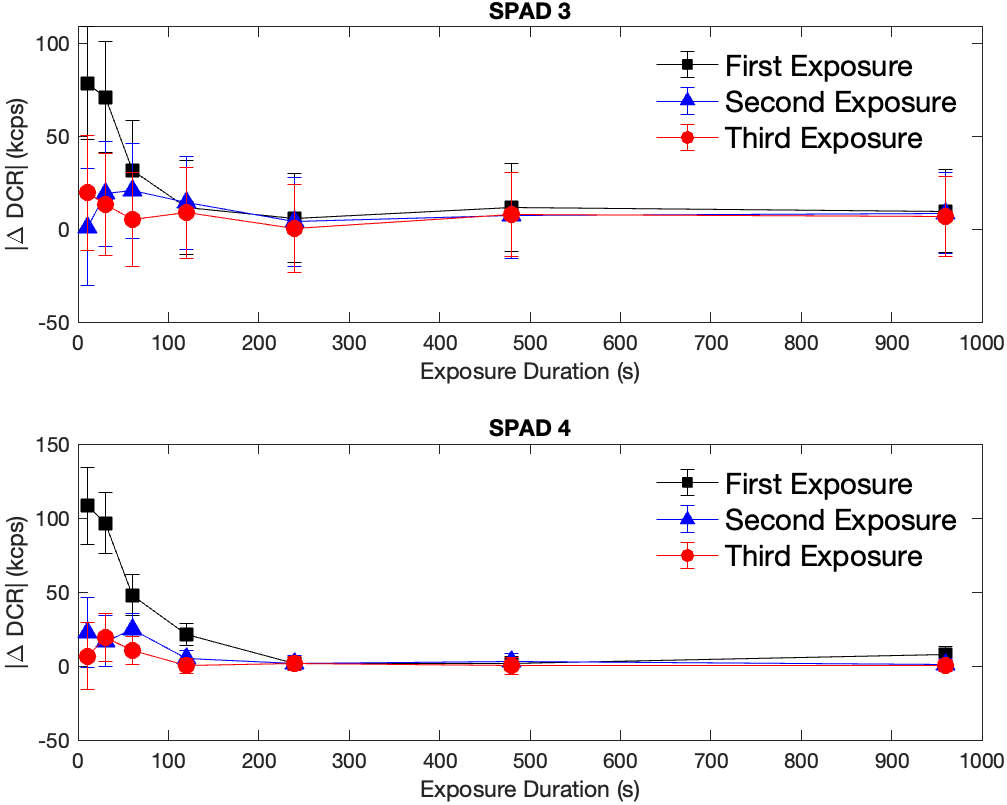}
	\caption{Absolute DCR change between subsequent exposures $\left|\Delta \text{DCR}\right| = \left|\text{DCR}(n+1) - \text{DCR}(n)\right|$ for SPAD~3 (top) and SPAD~4 (bottom). The largest $\left|\Delta \text{DCR}\right|$ is seen consistently after the first exposure at a new annealing duration (black curve).}
	\label{fig:triple-exposure-annealing}
\end{figure}

We did observe, however, that in the triple exposure protocol, the greatest change in DCR always occurred after the first of three rounds of annealing (\cref{fig:triple-exposure-annealing}). The behaviour is most evident for the shortest exposures, where the annealing duration is very short compared to the inter-annealing characterization time (seconds versus minutes). As the annealing time extends, this behaviour is less prominent. We propose that the earliest exposure of a new duration has access to previously inaccessible defects and neutralizes them, such that later exposures simply have fewer defects to anneal. This is also supported by the fact that a smaller change in DCR is observed later in the experiment, despite more energy being deposited per exposure. This result provides some guidance with respect to tactics of annealing in orbit: shorter exposures at a single energy could be applied until little change in DCR is observed, after which the energy can be increased, either by increasing the laser power or the exposure duration. 

\section{Discussion} 

\begin{table*}
\centering
\caption{Comparison of annealing studies}
\label{tab:annealing-study-comparison}
\resizebox{\textwidth}{!}{
\begin{tabular}{@{\extracolsep{1.8ex}}l@{}c@{}c@{}c@{}c@{}c@{}c@{}c@{}c@{}c@{}c@{}c@{}}
\hline\hline
Sample type & \makecell{Particle\\ type} & \makecell{Energy \\ (MeV)} & \makecell{Maximum\\ fluence\\ (particle/cm$^2$)} & \makecell{Displacement \\ damage dose \\ (MeV$\,$g$^{-1}$)} & DCIF & \makecell{Annealing\\ apparatus} & \makecell{Temperature~($\celsius$)\\ or optical\\ power~(W)} & Time (h) & DCRF & Additional notes & Reference \\
\hline
SLiK SPADs & Proton & $100$ & \num{2d10} & \num{5.95d7} & $1000$ & Laser & $1.8$~W & $0.25$ & $48$ & \makecell{Sampled previously\\ thermally annealed} & \makecell{Current\\ work} \\
SLiK SPADs & Proton & $100$ & \num{2d10} & \num{5.95d7} & $1000$ & TEC & $80$ & $1$ & $32$ & \makecell{Repeated cycles\\ of irradiation\\ and annealing} & \cite{dsouza2021} \\
SLiK SPADs & Proton & $100$ & \num{4d9} & \num{1.20d7} & $2500$ & \makecell{Oven,\\ TEC} & $80$, $100$ & $2$, $8$ & $6.66$ & & \cite{anisimova2017} \\
SLiK SPADs & Proton & $100$ & \num{4d9} & \num{1.20d7} & $43$ & Laser & $1.6$~W & $0.05$ & $41.7$ & \makecell{Samples previously\\ thermally annealed} & \cite{lim2017} \\
APD & Neutron & $1$ & \num{2d13} & \num{4.3d10} & $1000$ & Oven & RT to $60$ & \makecell{Up to\\ $120$~days} & $3.8$ & & \cite{baccaro1999} \\
APD & Proton & $51$ & \num{10d12} & \num{4.41d10} & $100$ & Oven & $100$ & $24$ & $2.32$ & & \cite{becker2003} \\
APD & Electron & $0.008$ & \num{10d15} & $0$ & $100$ & Oven & $226$ & $10$ & $33$, $100$ & & \cite{kawauchi2013} \\
\makecell[l]{PIN\\ photodiodes}& Proton & $60$ & \num{5.5d12} & \num{2.21d10} & $1000$ & None & RT & $6$~months & $2$ & \makecell{Illuminated with\\ $60$~W tungsten\\ light} & \cite{jimenez2007} \\
PM & Neutron & $14$ & \num{10d12} & \num{3.83d10} & $316$ & Oven & $250$ & \makecell{Up to\\ $5$~days} & $33$ & \makecell{TVAC, biased;\\ regain photon\\ number resolving\\ capability} & \cite{tsang2016} \\
PM & Neutron & $14$ & \num{10d10} & \num{3.38d8} & $316$ & Oven & $250$ & $3$ days & $33$ & TVAC, biased & \cite{tsang2018} \\
PM & Proton & $58$ & \num{1d8} & \num{4.1d5} & $100$ & \makecell{Cooler,\\ oven} & $-22$ to $49$ & $24$ & $2.5$ & \makecell{No effect\\ below RT} & \cite{deAngelis2023} \\
PM & Proton & $800$ & \num{4d11} & \num{5.63d8} & $10000$ & None & RT & $30$~days & $2$ & & \cite{bartlett2020} \\
\hline\hline
\end{tabular}
}
\end{table*}

\subsection{Comparison to thermal annealing}

\Cref{tab:annealing-study-comparison} shows our result in comparison with other reported radiation tests with similar silicon-based photon detectors and annealing. For ease of comparison, the displacement damage dose (DDD) was calculated using the online tool \textit{Screened-relativistic non-ionizing energy loss calculator}  \footnote{Note, the NIEL for electrons at the disclosed energy of the study ($8$~keV) should not contribute to DDD; however, the authors describe a sizeable increase in dark current, as well as a notable decrease in dark current after annealing. Therefore, the study was included in the comparison.}\cite{sr-niel}. For each study, the maximal increase in dark current or dark count rate above the pre-irradiation level, as well as the maximal reduction in the noise post-annealing is extracted, to estimate the dark current (count) increase factor (DCIF) or dark current (count) reduction factor (DCRF), respectively. 

The dominant method of annealing is either at room temperature (RT) or at elevated temperature in ovens. To our knowledge, Lim et al.\ \cite{lim2017} and the results disclosed in this article are the sole laser annealing studies. Jimenez et al.\ \cite{jimenez2007} illuminated a portion of their irradiated photodiodes while annealing at RT and found that there was an increased rate of recovery for the illuminated samples compared with the samples left unilluminated. Besides this instance, no other studies exposed irradiated silicon detectors to light during annealing. 

Comparison between traditional thermal annealing techniques and laser annealing highlights the rapidity of the latter: RT annealing requires multiple days or months, while raising the temperature via convection heating or TECs still requires several hours to achieve similar DCRF as a few minutes of laser annealing. In comparison to other studies, our results have the highest DCRF achieved in a much shorter time (omitting the low-energy electron study of \cite{kawauchi2013}, which incurred a negligible DDD). Moreover, the majority of the studies relied on ovens which would be infeasible for space applications. TECs are the closest alternative, but require significant current draw (a few amperes) for time periods on the order of an hour \cite{dsouza2021,anisimova2017}.

Outside the regime of image sensors, laser annealing has also been favored over conventional oven or furnace annealing methods in semiconductor fabrication. Some example applications utilizing laser annealing instead of thermal annealing are repairing the crystalline structure after ion implantation \cite{poate1982}, custom dopant profile design \cite{hill1981}, rapid crystal growth \cite{galvin1982}, formation of superalloys \cite{white1980}, as well as grain-size reduction \cite{gat1978}. In most cases, these applications use pulsed annealing, with the modularity of pulse length stated as the critical factor for determining the end result of the application \cite{white1983}.

\subsection{Proposed defect annealing mechanism}

As the exact mechanism by which defect annealing occurs remains unclear and why laser annealing appears to be quicker at achieving reduction in DCR in SPADs compared with traditional thermal annealing, we provide discussion on the possible ongoing physics, with the support from the field of semiconductor fabrication. 

The dark current and dark count rate are the result of collection of thermally-generated carriers, which can be either electrons collected at the positive node or holes collected at the negative node. In the case of the SPADs studied here, the signal is generated at the anode. A reduction in DCR or dark current is attributed to an increased mobility of defects, thereby reducing the quantity of generation centres in the depleted region. The mobility of defects is enhanced by the recombination processes occurring during the annealing exposure \cite{belykh1990,weeks1975}.

During proton irradiation, the bulk semiconductor is damaged, inducing vacancies and defect complexes, which can be sources of as well as traps for carriers \cite{holmes1993}. When the SPAD is unbiased, no additional carriers are introduced so the system is in equilibrium: recombination processes occur, but at the same rate as generation. However, when the SPAD is exposed to the annealing laser, additional carriers are introduced via the processes of absorption and ionization. Some of these carriers recombine through radiative mechanisms, with photons being released as the system relaxes; other carriers recombine through non-radiative processes, resulting in energy transferred into phonons, increasing the lattice vibrations and ultimately the lattice temperature \cite{white1983,weeks1975,baeri1996,bell1979}. It has been shown that high intensity laser pulses on the order of nanosecond duration with the appropriate wavelength may deposit enough energy to increase the temperature above melting \cite{white1983}. On the other hand, there have also been studies that measure temperatures well below the melting point despite the sample undergoing characteristic changes associated with melting \cite{lo1980,foti1978}. It is believed that these inconsistencies in phase behaviour are due to the localized rapid heating and cooling resulting in novel structures with unique properties \cite{white1983}. 
 
While our study does not use pulsing, the monotonic deposition of energy must induce structural changes since we observe a permanent reduction in dark count rate. This reduction could be a manifestation of a mechanism proposed by Weeks et al.\ \cite{weeks1975}, who suggest that defect reaction rates are increased as a by-product of recombination processes occurring during the introduction and subsequent capture of injected carriers. Their proposed rate equations are in good agreement with the experimental results of Kimerling and Lang \cite{KL1975}, who observed higher annealing rates in electron-irradiated GaAs after forward biasing (carrier injection). Moreover, Weeks et al.\ \cite{weeks1975} show that the activation energy for this recombination-enhanced defect annealing process is lower than that required in thermal equilibrium. This, coupled with the possibility of rapid-onset local heating, could explain why optically-stimulated carrier injection yields a quicker defect annealing rate compared with traditional thermal annealing through convective heating. 

Finally, the SPAD is biased after annealing exposure and only those carriers which were not annihilated through the aforementioned annealing process contribute to the measured dark count rate.

We now examine the plausibility of the recombination-enhanced annealing mechanism in the case of optical injection of carriers. When there are more carriers present compared with the thermal equilibrium concentration, recombination processes will occur in order to return to equilibrium conditions. For significant annealing to take place as result of non-radiated energy loss during recombination, $np \gg n_i^2$, where $n$ and $p$ are the negative and positive charge carrier concentrations and $n_i$ is the intrinsic carrier concentration, which is $n_i^2 \approx 10^{20}$~cm$^{-6}$ in silicon \cite{hu2010}. The injected carrier concentration is derived for an optical power of $1$~W, which is the lowest optical power yielding a measurable decrease in DCR. For our annealing laser wavelength of $808$~nm and estimated illuminated area of $90~\micro$m radius, $1$~W corresponds to a photon flux of \num{1.60d22}~cm$^{-2}$s$^{-1}$. According to the Beer-Lambert Law, the intensity $I$ of the photon flux at position $z$ of a substance will decrease from its initial intensity $I_0$ according to the exponential $I(z) = I_0 e^{-\alpha z}$, where $\alpha$ is the absorption decay coefficient. For $808$~nm photons impinging on intrinsic silicon, $\alpha$ is $790$~cm$^{-1}$, and the penetration depth ($\alpha^{-1}$) is $\approx 12~\micro$m \cite{green1995}. With such a penetration depth, we expect the majority of the photons to be absorbed by around $33~\micro$m. According to Dautet et al.\ \cite{dautet1993}, the typical thickness for SLiK SPADs such as those presented here is $20$--$30~\micro$m. We assume then that the full thickness of the active region will contribute to carrier generation. Accounting for fiber losses and using the manufacturer data to estimate a lower bound on the coupling and photon detection efficiencies ($90\%$ and $60\%$, respectively), we conservatively estimate that $55\%$ of the photons are absorbed in the active region. The local electron-hole pair generation rate per unit volume is then calculated to be \num{1.55d23}~cm$^{-3}$~s$^{-1}$. Accounting for an average carrier lifetime in similar devices to be on the order of $\micro$s, we estimate the steady-state carrier density as \num{1.55d17}~cm$^{-3}$.

Comparing to $n_i = 10^{10}$~cm$^{-3}$, we see that, indeed, the number of injected carriers is much greater than the intrinsic population by several orders of magnitude. This supports our theory that significant recombination should be occurring when the annealing laser is injecting carriers, and these recombination processes could be aiding in the removal of generation centres.

\subsection{Potential adverse effects from laser annealing} 

The observed increase in DCR after annealing in the case of SPAD~4 is hypothesized to be a result of addition of defects by the annealing laser. Direct ionization dose is not believed to be the cause of this, as examples in literature have not seen major changes in SPAD characteristics for much more energetic wavelengths (e.g.,\ under $^{60}$Co irradiation), and when there are changes they are attributed to displacement damage from secondary Compton electrons \cite{becker2003,baccaro1999,tan2013,marisaldi2011,moscatelli2013}. We believe the damage is more likely due to thermal degradation of the bulk silicon, which has been observed for similar cw laser powers on silicon photodiodes \cite{beechem2013}, as well as in avalanche photodetectors, albeit with short-pulsed lasers at longer wavelengths \cite{watkins1990,watkins1990book,acharekar1988}. The results of Beechem et al.\ \cite{beechem2013} are most compelling, in that a clear increase in dark current was observed with laser powers $>2$~W despite the temperature of the silicon never reaching the melting point of $1414~\celsius$. Simulation in the same study revealed that decreases in $N_a$, the acceptor atom concentration, and $\tau_0$, the minority carrier lifetime, yielded rises in dark current comparable to those observed in the exposed samples. As such, we conclude that it is conceivable that too powerful laser exposure can alter the defect concentration such that the dark count rate \textit{increases.} 

While constraints on safe levels of laser exposure are yet to be established, our study indicates that short exposures on the order of tens of seconds already diminish the effects of displacement damage with no reportable adverse consequences.

\section{Conclusion}

We study how laser annealing can be a very impactful method of reducing dark counts, and can increase the signal-to-noise ratio for single-photon signals. We demonstrate several annealing protocols---chosen to be feasible in-orbit---in a vacuum environment, on four irradiated SPADs within a detector module developed for a CubeSat mission. Our test and comparison of different annealing protocols appears to show that the annealing power is more critical than laser annealing duration, as we find maximal reduction in DCR occurring after short bursts of high power exposure of $1.8$~W, rather than prolonged annealing at a lower power of $1.2$ or $1.5$~W. As such, periodic, short ($<30$~s) annealing exposures can be strategic for managing the daily accumulation of dark counts within the constrained power budget of a small satellite. We observe that DCR plateaus after about $100$~min of accumulated annealing at a single power, although subsequent exposure to a higher power can yield further improvement. Our results suggest that in-orbit annealing laser circuitry should be equipped to handle a range of output optical powers at least up to $2$~W that can be adjusted remotely based on real-time DCR and NCR measurements. The improvement of detector operation from laser annealing is evident as the sensitivity of a detector at a fixed bias is increased, and the saturation effects are mitigated. Our demonstration is the first to achieve high-power annealing in fiber-coupled devices, the first to operate under a simulated space environment, and shows that the SPAD active area and fiber coupling optics can tolerate very high intensity and prolonged (several hours) stimulation.

\section*{Acknowledgements}

We acknowledge Paul Kwiat and Michael Lembeck of University of Illinois at Urbana-Champagne for their input on the design of the annealing apparatus of CAPSat, as well as Jin Lim, Nachiket Sherlekar and Jean-Phillipe Bourgoin for their helpful discussion. This project received funding support from the National Science and Engineering Council of Canada (NSERC), the Canadian Space Agency through their FAST grant, the Canadian Foundation for Innovation (CFI), and the Ontario Research Foundation (ORF).

\bibliographystyle{IEEEtran}
\bibliography{bibliofile}

\end{document}